\documentclass[pra,twocolumn,twoside,aps]{revtex4-1}

\usepackage{amsmath, amssymb}
\usepackage{bm}
\usepackage{natbib}
\usepackage{color}

\usepackage{graphicx}
\usepackage{float}

\begin{document}
\title{Nonlinear Zel'dovich effect: Parametric amplification from medium rotation}
\author{Daniele Faccio$^{1,2}$ and Ewan M. Wright$^{2,1}$}
\affiliation{$^1$Institute of Photonics and Quantum Sciences, Heriot-Watt University, Edinburgh EH14 4AS, UK}
\affiliation{$^2$College of Optical Sciences, University of Arizona, Tucson, Arizona 85721, USA}
\begin{abstract}
The interaction of light with rotating media has attracted recent interest for both fundamental and applied studies including rotational Doppler shift measurements. It is also possible to obtain amplification through the scattering of light with orbital angular momentum from a rotating and absorbing cylinder, as proposed by Zel'dovich more than 40 years ago. This amplification mechanism has never been observed experimentally yet has connections to other fields such as Penrose superradiance in rotating black holes.  Here we propose a nonlinear optics system whereby incident light carrying orbital angular momentum drives parametric interaction in a rotating medium. The crystal rotation is shown to take the phase-mismatched parametric interaction with negligible energy exchange at zero rotation to amplification for sufficiently large rotation rates. The amplification is shown to result from breaking of anti-PT symmetry induced by the medium rotation.
\end{abstract}

\maketitle
\noindent
{\bf Introduction:} The interaction of vortex light beams carrying orbital angular momentum (OAM) with rotating media has been shown to lead to a series of {novel fundamental} phenomena {and applications}. Some such as the rotational Doppler shift have an analogue for non-rotating light/media \cite{Birula,Padgett,SpeLavPad14} whilst others allow new effects such as the creation of effective magnetic fields for light \cite{magnetic}.  A recent study of second-harmonic generation in a rotating crystal showed the existence of an unexpected nonlinear analogue of the rotational Doppler effect, i.e. a frequency shift imparted upon a beam with OAM from a rotating crystal \cite{Zhang}.\\
\indent Zel'dovich first described the situation in which a material cylinder that is an absorber of incident radiation while at rest, could nonetheless amplify incident light waves carrying optical angular momentum  if the cylinder was rotating at a high enough frequency $\Omega$ around its axis \cite{Zel,Zel2}. In this way energy of rotation of the medium can be transferred to the light field, a result whose generalization encompasses the extraction of energy from rotating black holes or stars {\cite{Penrose,bomb,Cardoso1,Cardoso2}}.  An elementary picture of how the Zel'dovich effect arises may be garnered from considering a cylinder made up of two level atoms, and a probe field of frequency $\omega_1$ carrying OAM with winding number $\bar m$.  In this case the linear susceptibility of the medium as calculated in the reference frame rotating at frequency $\Omega$ may be written as the sum of two Lorentzians \cite{TLA}
\begin{equation}
\chi^{(1)}(\omega') = \left [ {N|\mu|^2/\epsilon_0\hbar\over \omega_0-\omega' - i\Gamma/2} + {N|\mu|^2/\epsilon_0\hbar\over \omega_0+\omega' + i\Gamma/2}\right ]  , 
\end{equation}
where $N$ is the number density of atoms, $\mu$ the dipole matrix element between the two levels, $\omega_0$ being the transition frequency, $\Gamma$ is the population decay rate of the upper level, and $\omega'=(\omega_1-\bar m\Omega)$ accounts for the rotational Doppler effect \cite{Padgett}.  For the non-rotating case and $\omega'=\omega_1\simeq \omega_0$ for near-resonant conditions, the second Lorentzian in the square brackets may be neglected on the basis that it is non-resonant, and this yields a net absorption.  In contrast for a large enough rotation rate $\omega'$ can become negative and the second Lorentzian can become resonant and dominant.  In this case a net gain arises since the second Lorentzian has the opposite sign of the upper level decay rate.  Gain then becomes a possibility for $\omega'<0$, i.e.
\begin{equation}\label{mOm}
\bar m\Omega > \omega_1 ,
\end{equation}
which is the condition commonly quoted for observing the Zel'dovich effect \cite{Zel,Zel2,BekSch98,MIT}. \\ 
\indent Here we consider a nonlinear optics realization of the Zel'dovich effect that emerges from three-wave mixing of ring-shaped vortex beams in a rotating second-order nonlinear crystal.  
We find that a light beam carrying OAM can experience parametric amplification under a condition on the crystal rotation rate akin to Eq.~(\ref{mOm}).  The key physics is that the rotation modifies the phase-matching of the nonlinear interaction, which is phase-mismatched at zero-rotation, and triggers parametric amplification for sufficient rotation.  This amplification is shown to result from breaking of anti-PT symmetry induced by the rotation.\\
\noindent
{\bf Basic geometry and equations:} Our basic model involves propagation along the optic axis in a nonlinear uniaxial optical crystal: As a concrete example we choose a crystal of point symmetry $32$ as described in Ref. \cite{BeyRabin67}, but the approach applies to other point symmetries such as $3m$.  In our model of parametric amplification a signal field at the fundamental frequency $\omega_1$ is incident on the second-order nonlinear crystal along with a pump field at the second-harmonic (SH) frequency $\omega_2=2\omega_1$.  In this case the nonlinear parametric interaction can generate an idler field that is also at the fundamental frequency $\omega_3=(\omega_2-\omega_1)=\omega_1$.   For this geometry it is known that if the fundamental field is circularly polarized (same handedness for both signal and idler) the SH field has the opposite handedness \cite{BeyRabin67}.  Denoting the complex amplitude of the circularly polarized fundamental field and of the oppositely handed circularly polarized SH field as $A_1(x,y,z)$ and $A_2(x,y,z)$, respectively, the slowly-varying envelope equations used in Ref. \cite{BeyRabin67} for the fields take the form (for more detail see Sec. I of the appendix)
\begin{eqnarray}\label{Eqs2}
\frac{\partial A_1}{\partial z}&=&{i\over 2k_1} \nabla_\perp^2 A_1 + i\eta A_2A_{1}^*e^{-i\Delta k z}  ,\nonumber \\
\frac{\partial A_2}{\partial z}&=&{i\over 4k_1} \nabla_\perp^2 A_2 + i\eta A_1^2 e^{i\Delta k z}  ,
\end{eqnarray}
where $k_j=n_j\omega_j/c$, $n_j=n_o(\omega_j)$ is the ordinary refractive-index at the selected frequency, $\nabla_\perp^2$ is the transverse Laplacian describing diffraction, $\eta=2d_{eff}\omega_1/n_1c$ with $d_{11}$ the second-order nonlinear coefficient, $\Delta k=2k_1-k_2$, and we used $k_2\approx 2k_1$ in the SH diffraction term. Equations (\ref{Eqs2}) are the basis for our subsequent development and coincide in form with those given by Boyd \cite{Boyd} and also used in Ref. \cite{LowRogFac14}.

\noindent
{\bf Rotating frame equations:} Our goal is to investigate the parametric interaction between the fields in a frame rotating at frequency $\Omega$ around the optic axis, Eqs. (\ref{Eqs2}) being in the lab frame.  We note that the nonlinear terms in these equations are invariant with respect to rotation due to the choice of propagation along the optic axis and the use of circular polarization states.  To proceed we state the field equations in the rotating frame:
\begin{eqnarray}\label{EqsRot}
\frac{\partial A_1}{\partial z}&=&{i\over 2k_1} \nabla_\perp^2 A_1 \underbrace{-k_1\left ({\Omega\over \omega_1} \right ){\partial A_1\over\partial \phi}} + i\eta A_2A_{1}^*e^{-i\Delta k z}  ,\nonumber \\
\frac{\partial A_2}{\partial z}&=&{i\over 4k_1} \nabla_\perp^2 A_2 \underbrace{-k_2\left ({\Omega\over \omega_2} \right ){\partial A_2\over\partial \phi}}+ i\eta A_1^2 e^{i\Delta k z}  ,
\end{eqnarray}
with $\phi$ the azimuthal angle in cylindrical coordinates $(\rho,\phi,z)$.  The underbraced terms represent the effect of transforming to the rotating frame and may be understood as follows:  If we consider either field with winding number $\ell$, and associated azimuthal variation $e^{i\ell\phi}$, the underbraced terms may be written generically as
$$
-k\left ({\Omega\over \omega} \right ){\partial A\over\partial \phi} \equiv -ik\left ({\ell\Omega\over \omega} \right )A=i\delta kA  ,
$$
where we have dropped the subscript $j=1,2$ for simplicity.  Using this result in combination with Eqs.~(\ref{EqsRot}), we identify the fractional change in the longitudinal wavenumber as $\delta k=-{\ell\Omega\over\omega}k$ for beams carrying OAM, in agreement with Ref.~\cite{SpeLavPad14}.   Then the longitudinal wavenumber in the rotating frame is $k'=k\left ( 1 - {\ell\Omega\over\omega} \right )$, and there is a concomitant rotational Doppler shifted frequency $\omega'=\omega\left ( 1 - {\ell\Omega\over\omega} \right )$.  The underbraced terms in Eqs.~(\ref{EqsRot}) therefore account for the rotational Doppler effect in the rotating frame.

\noindent
{\bf Perfect optical vortices:} Our proposal for the nonlinear Zel'dovich effect (NLZE) involves the parametric interaction between weak signal and idler fields in the presence of a strong SH pump field.  Since the signal and idler fields are both at the fundamental frequency they must be distinguished in some other way. To develop the ideas and have an analytic theory we consider the case that all interacting fields are perfect optical vortices (POVs) \cite{OstRicArr13,CheMalAri13} with different helical phase-front winding number $m$.  POVs are ring-shaped beams whose radius $R$ is independent of winding number and the same for all interacting fields.  As shown in Sec. II of the appendix, for POVs of width $W$, $R>>W>>\lambda$, the slowly varying electric field envelope for a POV around the peak of the ring may be written as
\begin{equation}
A(\rho=R,\phi,z) = a(z)e^{im\phi}e^{-{iz\over 2k}{m^2\over R^2} -ikz{m\Omega\over \omega}}  .
\end{equation}
In the second exponential on the right-hand-side the first term describes the reduction in the z-component of the wavevector due to the ray skewing associated with the beam OAM \cite{DhoSimPad96,RogHeiWri13}, and the second term accounts for the rotational Doppler effect.

\noindent
{\bf Parametric interaction of POVs:}
To proceed we assume that the pump $(j=2)$ field is much stronger than the signal $(j=1)$ field.  Then the parametric amplification process, which produces one signal and one idler photon from one pump photon, generates an idler field $(j=3)$ that has winding number $m_3=m_2-m_1$.  Assuming all fields are described by POVs we then write the slowly varying electric fields for the fundamental and second harmonic fields, with $\rho=R$, as
\begin{eqnarray}\label{Aexpan}
A_1(\phi,z) &=& a_1(z)e^{im_1\phi}e^{-{iz\over 2k_1}{m_1^2\over R^2} -ik_1z{m_1\Omega\over \omega_1}}+ \nonumber \\  
&+& a_3(z)e^{im_3\phi}e^{-{iz\over 2k_1}{m_3^2\over R^2} - ik_1z{m_3\Omega\over \omega_1}} , \nonumber \\
A_2(\phi,z) &=& a_2e^{im_2\phi}e^{-{iz\over 4k_1}{m_2^2\over R^2} - ik_2 z{m_2\Omega\over \omega_2}},
\end{eqnarray}
with $a_2$ independent of $z$ in the undepleted pump beam approximation, and $a_3(0)=0$ with no idler present at the input.  Here we have set $k_3=k_1$ since the signal and idler have the same frequency and experience the same refractive-index. In Sec. III of the appendix we show that using the fields in Eqs. (\ref{Aexpan}) along with the propagation Eqs. (\ref{EqsRot}) yields the linearized signal-idler equations in the rotating frame
\begin{equation}\label{a1a3}
{da_1\over dz}= i(\eta a_2) a_3^*e^{i\kappa z}, \quad
{da_3\over dz}= i(\eta a_2) a_1^*e^{i\kappa z}  ,
\end{equation}
where the OAM dependent wavevector mismatch is
\begin{eqnarray}\label{kappa}
\kappa &=& -\Delta k + {\Omega m_2\over c} \left ( n_1 - n_2\right )  + {(m_1-m_2/2)^2\over k_1R^2}  \nonumber \\
&\approx & {2\over c} \left ( \omega_1 -\bar m \Omega \right ) (n_2-n_1) + {(m_1-m_2/2)^2\over k_1R^2}  .
\end{eqnarray}
Here $\bar m=m_2/2=(m_1+m_3)/2$ may be viewed as the mean winding number of the combined signal and idler fields.  These equations may be solved for the fields at the output of the crystal of length $L$ \cite{Boyd,LowRogFac14}.  The detailed expressions are given in Sec. IV of the appendix, with the final result that the net gain for the fundamental field (combined signal and idler output power over input signal power) may be expressed as  
\begin{equation}\label{G}
G=\left | \cosh(gL) - {i\kappa\over 2g}\sinh(gL)\right |^2 +  \left | {\eta a_2\over g}\sinh(gL)\right |^2
\end{equation}
and the signal gain (output over input signal power) is
\begin{equation}\label{Gs}
G_s=\frac{P_s(L)}{P_{sig}}=\left | \cosh(gL) - {i\kappa\over 2g}\sinh(gL)\right |^2  .
\end{equation}
Here $g = \sqrt{\beta I_p -\kappa^2/4}$ is the growth rate if the argument of the square root is positive.
\begin{figure}[t!]
\includegraphics[width=8cm]{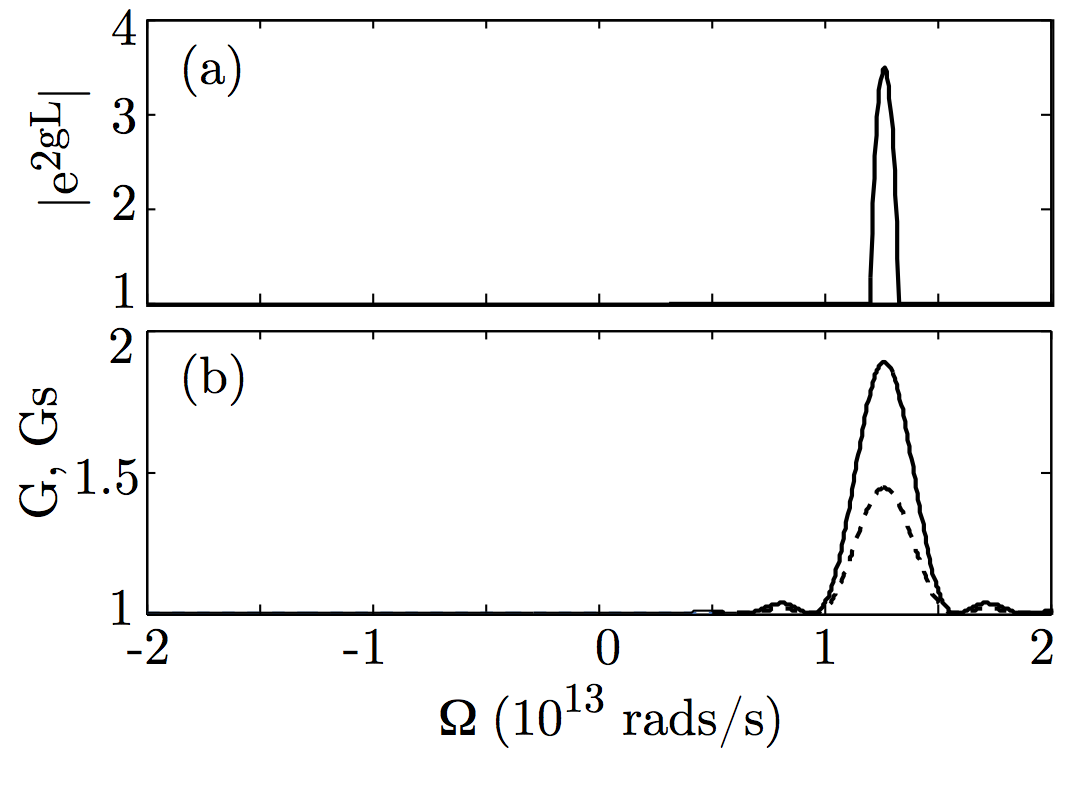}
\caption{(a) Parametric gain factor $|\exp(2gL)|$ over the medium length versus the rotation rate $\Omega$, and 
(b) the signal gain $G_s$ (dash line) and net gain $G$ (solid line) both as functions of the rotation rate $\Omega$.  For these calculations $m_1=149$ and $m_2=300$. }\label{Fig1}
\end{figure}

\noindent
{\bf Nonlinear Zel'dovich effect:} From the growth rate $g$ above it follows that parametric amplification arises for $\beta I_p > \kappa^2/4$, the growth rate being zero for $\beta I_p \le \kappa^2/4$.  Consider a situation in which for zero rotation, $\Omega=0$, the growth rate is zero, $\beta I_p < \kappa^2/4$.  If we consider normal dispersion so that $n_2> n_1$ and $\bar m\Omega > 0 $, then according to Eq.~(\ref{kappa}) with a large enough rotation rate a non-zero growth rate can arise due to rotation, that is due to the effect of non-zero $\Omega$ reducing $\kappa^2$.  For illustration, if we neglect the second term on the bottom line of Eq. (\ref{kappa}), based on taking the limit $k_1R >> 1$, the condition for $\kappa=0$ and maximal growth rate, becomes
\begin{equation}\label{peak}
 \omega_1 = \bar m\Omega .
 \end{equation}
This expression coincides with the boundary between loss and gain in Eq.~(\ref{mOm}) found by Zel'dovich \cite{Zel,Zel2}.  In our case the probe is composed of both signal and idler fields so the mean winding number $\bar m$ appears, and Eq.~(\ref{peak}) corresponds to the peak parametric amplification.

Figure \ref{Fig1} shows an example of the predicted parametric amplification at $\lambda_1=1~\mu$m arising from rotation for a crystal of length $L=2$ mm, nonlinear coefficient $d_{eff}=0.83$ pm/V, $n_1=1.6, (n_2-n_1)=10^{-3}$, pump intensity $I_p=2$ GW/cm$^2$, and a ring radius $R=32~\mu$m.  Furthermore, we set $m_2=300$ and take $m_1=149$ giving $m_3=151$.  This choice implies that $(m_1-m_2/2)=1$ is minimized in the last term in Eq. (\ref{kappa}), while keeping $m_{1,3}$ distinct and $m_2=2\bar m$ large.  Figure \ref{Fig1}(a) shows the predicted parametric gain factor $|\exp(2gL)|$ over the medium length versus the rotation rate $\Omega$.  This plot is compatible with our discussion above of the NLZE:  First, since $\bar m>0$ parametric amplification is possible only for $\Omega>0$, so the medium rotation and probe OAM must be co-rotating for gain, in agreement with the LZE.   In addition we find the peak gain for $\Omega_p=\omega_1/\bar m=1.2\times 10^{13}$ rads$^{-1}$.  The bandwidth of the parametric amplification may be estimated using the condition for growth $-\sqrt{\beta I_p}<\kappa/2 < \sqrt{\beta I_p}$.  Then using the previous approximation $k_1 R >> 1$ we obtain $\delta\Omega \approx {2 c\sqrt{\beta I_p}/[ (n_2-n_1)\bar m]}$. For the parameters used $\delta\Omega=0.1\times 10^{13}$ rads$^{-1}$ in agreement with Fig. \ref{Fig1}(a).

Note that the factor $|\exp(2gL)|$ only shows the material gain. In experiments aimed at revealing the NLZE, one would inject a signal field and measure the net and signal gains given in Eqs.~(\ref{G}) and (\ref{Gs}), respectively.  Figure \ref{Fig1}(b) shows the signal gain versus $\Omega$ (dash line) and the net gain (solid line), signal plus idler.  Thus, whether the signal alone is detected or both the signal and idler, clear amplification is observed over a range of positive rotation rates.  The full-width $\Delta\Omega$ for the net and signal gains is larger than the parametric gain in Fig.~\ref{Fig1}(a), meaning that the fields can still exchange energy even outside of the gain region, and may be estimated by requiring $\kappa L=\pm\pi$ at the edges for the phase-mismatch to diminish the gain.  Using, as above, the approximation $k_1R >> 1$  yields $\Delta\Omega \approx {\pi c/[ (n_2-n_1)\bar m L]} $. For the parameters used here this yields $\Delta\Omega=0.3\times 10^{13}$ rads$^{-1}$ in reasonable agreement with Fig.~\ref{Fig1}(b).

\begin{figure}[t!]
\includegraphics[width=8cm]{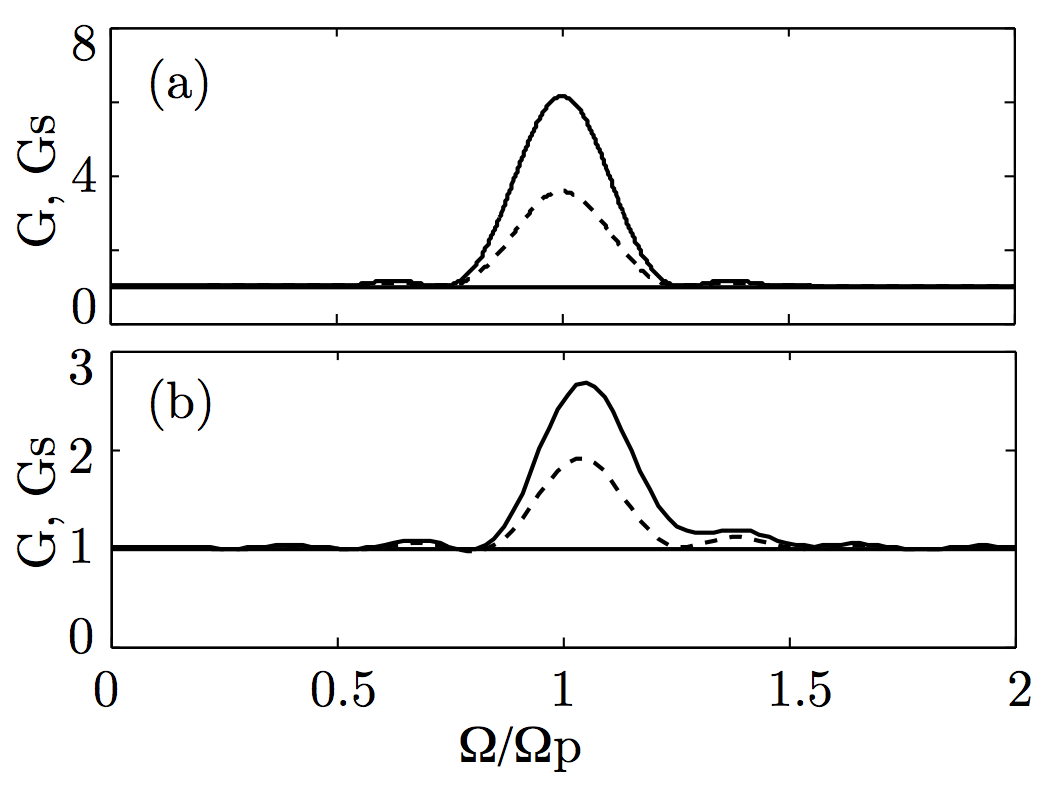}
\caption{Signal gain $G_s$ (dashed line) and net gain $G$ (solid line) both as functions of the scaled rotation rate $\Omega/\Omega_p$ using (a) the analytic theory and (b) the BPM.  For these calculations $m_1=8$ and $m_2=17$, all other parameters being the same as for Fig. \ref{Fig1}}\label{Fig2}
\end{figure}

\noindent
{\bf Numerical simulations:}
We performed beam propagation method (BPM) simulations in order to verify our results, independently of the approximations employed above.
We first note that the frequency width $\Delta\Omega$ given above, when normalized to the peak rotation rate $\Omega_p=\omega_1/\bar m$ becomes independent of the probe winding number $\bar m$.  From the perspective of comparing with BPM simulations it is therefore useful to look at the probe gain versus scaled rotation rate $(\Omega/\Omega_p)$.  This is particularly the case since including large field winding numbers in the BPM is computationally challenging.

Figure~\ref{Fig2} shows the results for the gain as a function of scaled rotation rate $(\Omega/\Omega_p)$ using (a) the analytic theory and (b) the BPM based on the propagation Eqs.~(\ref{EqsRot}) (solid lines are the net gain $G$ for the fundamental and dashed lines are the gain $G_s$ for the signal alone).  For these calculations $m_1=8$ and $m_2=17$, and the BPM simulation is performed with ring beams of radius $R=43~\mu$m as described in Ref. \cite{LowRogFac14}.  The fact that the analytic theory yields higher gains is not surprising given that it is based solely on the peak of the ring where the maximum gain appears, whereas the BPM includes the distribution of intensities in the fields.  However, the overall qualitative agreement between the BPM and analytic theory verifies the ideas and theory underlying the latter.

\noindent
{\bf Breaking of anti-PT symmetry:} Using the change of variables $a_{1,3}(z)=b_{1,3}(z)e^{i\kappa z/2}$, Eqs. (\ref{a1a3}) may be written in the matrix form
\begin{equation}
i{\partial\over\partial z}
\left (
\begin{array}{c}
b_1 \\ b_3^*
\end{array}
\right ) = U
\left (
\begin{array}{c}
b_1 \\ b_3^*
\end{array}
\right ) 
=\left (
\begin{array}{cc}
\kappa/2  -\eta a_2 \\
\eta a_2^*  -\kappa/2
\end{array}
\right )
\left (
\begin{array}{c}
b_1 \\ b_3^*
\end{array}
\right ) ,
\end{equation}
with interaction operator $U$.  We may view this as analogous to a two-state quantum system ($\hbar=1$) with $z$ playing the role of time ($T$) and $U$ the Hamiltonian, with the caveat that the Hamiltonian is not Hermitian in this case.  The energy eigenvalues of $U$ are given by $E=\pm \sqrt{\kappa^2/4-\beta I_p}$, where we used $\beta I_p=\eta^2|a_2|^2$, which are either both real or both imaginary.  This is reminiscent of the class of Hamiltonians that are non-Hermitian but can display parity-time (PT) symmetry and yield real eigenvalues \cite{BenBot98,Guo09,Christo}. More specifically, following Bender {\it et. al} \cite{BenBerMan02} the combined action of the parity operator $P$, which interchanges $1\leftrightarrow 3$, and the time-reversal operator $T$, which takes the complex conjugate, on the interaction operator yields $[PTU]_{\mu\nu}=U_{\nu\mu}^*$, with $\mu,\nu=\pm 1$ and the identifications $+1\equiv 1$, $-1\equiv 3$.  Then for the case with $|\kappa/2|\ge \sqrt{\beta I_p}$ with real eigenvalues we find $PTU=-U$, the real eigenvalues giving rise to a {\it phase-conjugate} coupling between the basis states with concomitant oscillatory dynamics. In this case the interaction operator displays {\it anti-PT symmetry} as recently revealed for parametric interactions in nonlinear optics \cite{Ant15}. In contrast, for the case $|\kappa/2|< \sqrt{\beta I_p}$ with imaginary eigenvalues $PTU=U$, and the system displays PT symmetry, or broken anti-PT symmetry. In this case the imaginary eigenvalues $E =\pm ig$ give rise to {\it parametric gain and loss}. We note that phase-conjugate coupling can also produce a net gain of an incident signal via energy exchange, and this underpins why the signal gain in Fig.~\ref{Fig1}(b) can occur over a full width $\Delta\Omega$ that is larger than $\delta\Omega$ in Fig.~\ref{Fig1}(a) for strict parametric amplification.  In our case the transition from unbroken to broken anti-PT symmetry is accomplished by rotating the nonlinear medium. Physically, for large enough rotation rates the peak of the POV ring beams acts as an ergoregion from which energy can be extracted from the rotational energy of the medium, that must be replenished to maintain the rotation, in the form of amplification of the probe beam.  In related earlier work Silveirinha \cite{Silv14} described spontaneous PT symmetry breaking as the result of linear motion of a third-order nonlinear medium, with concomitant modulation instability and amplification.  The role of PT symmetry in wave instabilities in a cavity with rotating walls was discussed in Ref. \cite{LanSil16}, this system having intimate connections with the linear Zel'dovich effect.\\
{\bf Conclusions:} Parametric interaction in a rotating crystal arises due to a `nonlinear' Zel'dovich effect whereby the rotational energy of the transparent crystal triggers parametric amplification of light signals. In the linear Zel'dovich effect the amplification arises from the rotational Doppler effect changing the resonance properties of the medium, whereas here the amplification arises from rotation-induced changes in phase-matching.   
As for the linear Zel'dovich effect the $\bar m\Omega=\omega_1$ condition leads to rotation rates of the order of THz even for $\bar m=1000$.  Lower rotations are expected by examining other forms of medium nonlinearity, for example stimulated scattering.  In this case the rotational Doppler shift could be used to change an incident field tuned to the anti-Stokes resonance at zero rotation, which experiences loss, into a Stokes wave with accompanying gain for sufficient rotation.  For Brillouin scattering the required rotation frequency is related to the Brillouin frequency shift (i.e. the frequency of the medium phonons, of the order of 1-0.1 GHz \cite{Brill}) as opposed to the optical frequency. This could bring the overall rotation frequencies towards the experimentally accessible MHz regime \cite{kishan}, although more detailed modeling will be required in order to quantitively verify this prediction.\\
{Our results extend ongoing studies of the interaction of matter with light possessing OAM. For example, OAM may modify the microscopic interaction symmetry and the selection rules with a single atom \cite{Davis1,Davis2,Ncomms}. Our work shows that beyond this, rotation of the medium may lead to a breaking of the macroscopic parity-time symmetry of the interaction that results in amplification of the optical beam at the expense of the medium rotation. Observing this amplification would not only be of importance for our understanding of fundamental phenomena but could lead to applications in quantum processing (through amplification of quantum vacuum states) with potential extensions also to plasmonics \cite{molina} or slow light systems that may further enhance the interaction \cite{boyd1,boyd2}.
}\\
\noindent
{\bf {Acknolwedgements.}} D. F. acknowledges financial support from the European Research Council under the European Union Seventh Framework Programme (FP/2007-2013)/ERC GA 306559 and EPSRC (UK, Grant No. EP/M009122/1).

\section{Appendix: Basic geometry and equations}
The basic model involves propagation along the optic axis in a nonlinear uniaxial optical crystal of point symmetry $32$ as described in Ref. [12].  A signal field at the fundamental frequency $\omega_1$ is incident on the second-order nonlinear crystal along with a pump field at the second-harmonic (SH) frequency $\omega_2=2\omega_1$, and the nonlinear parametric interaction generates an idler field that is also at the fundamental frequency $\omega_3=(\omega_2-\omega_1)=\omega_1$.   For this geometry it is known that if the fundamental field is circularly polarized (same handedness for both signal and idler) the SH field has the opposite handedness [12].  Then denoting the complex amplitudes of the right-handed $(-)$ and left-handed $(+)$ circularly polarized fundamental and SH fields as $E_{\omega_1}^\pm$ and $E_{\omega_2}^\pm$, respectively, the slowly-varying envelope equations for the fields become 
\begin{eqnarray}\label{Eqs1}
{\partial E_{\omega_1}^\pm\over \partial z } &=& {i\over 2k_1}\nabla_\perp^2 E_{\omega_1}^\pm
+{i\pi d_{11}\over k_1} \left ({\omega\over c} \right )^2 (E_{\omega_1}^\pm)^* E_{\omega_2}^\mp e^{i\Delta k_{SHG}z} ,\nonumber \\
{\partial E_{\omega_2}^\pm\over \partial z } &=& {i\over 2k_2}\nabla_\perp^2 E_{\omega_2}^\pm
+{i2\pi d_{11}\over k_2} \left ({\omega\over c} \right )^2 (E_{\omega_1}^\mp)^2 e^{-i\Delta k_{SHG}z}   ,
\end{eqnarray}
where $k_j=n_j\omega_j/c$, $n_j=n_o(\omega_j)$ being the ordinary refractive-index at the selected frequency, $\Delta k_{SHG}=k_2-2k_1$, $d_{11}$ is the relevant second-order nonlinear coefficient, and $\nabla_\perp^2$ is the transverse Laplacian describing diffraction.

We condense the notation with the definitions: $A_1(x,y,z)$ is the complex amplitude of the circularly polarized fundamental field and $A_2(x,y,z)$ is the complex amplitude of the oppositely handed circularly polarized SH field, $\eta=\pi d_{11}\omega_1/n_1c\approx \pi d_{11}\omega_1/n_2c$, and $\Delta k=2k_1-k_2=-\Delta k_{SHG}$.  With these definitions the paraxial wave equations for the fundamental $(j=1)$ and pump $(j=2)$ fields become
\begin{eqnarray}\label{Eqs2}
\frac{\partial A_1}{\partial z}&=&{i\over 2k_1} \nabla_\perp^2 A_1 + i\eta A_2A_{1}^*e^{-i\Delta k z}  ,\nonumber \\
\frac{\partial A_2}{\partial z}&=&{i\over 4k_1} \nabla_\perp^2 A_2 + i\eta A_1^2 e^{i\Delta k z}  ,
\end{eqnarray}
where we used $k_2\approx 2k_1$ in the SH diffraction term.  Adopting SI units, which entails replacing $\pi d_{11}=2d_{eff}$ and $\eta=2d_{eff}\omega_1/n_1c$, the output powers in the fundamental and SH fields for a medium of length $L$ can be expressed as
\begin{equation}
P_j(L) = 2\epsilon_0  n c \int_{-\infty}^\infty dx \int_{-\infty}^\infty dy |A_j(x,y,L)|^2, \quad j=1,2  .
\end{equation}
Equations (\ref{Eqs2}) are the basis for our subsequent development and coincide in form with those given by Boyd [13] and also used in Ref. [14].  This means that we can employ the ideas and analytical solutions given in those references.  We stress, however, that in these references the case of circularly polarized fields was not alluded to, so that making the connection between Eqs. (\ref{Eqs1}) and  Eqs. (\ref{Eqs2}) as done here for the point symmetry $32$ is a necessary step. 
\section{Appendix: Perfect optical vortices}
We present a representation of a monochromatic POV with frequency $\omega=2\pi c/\lambda$ and winding number $m$ propagating in a rotating medium of refractive-index $n$.  The POV has a ring shaped intensity profile of radius $R$ and width $W$, $R>>W>>\lambda$, along with a helical phase-front of winding number $m$. Assuming that the width $W$ of the POV is sufficiently narrow compared to the ring radius that we evaluate the properties of the beam around the peak of the ring.  Then for a POV with azimuthal variation $e^{im\phi}$ propagating along the z-axis, the corresponding spiraling wavevector may be written as [14]
\begin{eqnarray}\label{K}
\vec{K} &=& K_x\vec{e}_x + K_y\vec{e}_y + K_z\vec{e}_z  \nonumber\\
&=&  {m\over R}\cos(\phi)\vec{e}_x + {m\over R}\sin(\phi)\vec{e}_y + K_z\vec{e}_z  .
\end{eqnarray}
By demanding that $K=k'=k\left ( 1 - {m\Omega\over\omega} \right )$ in the rotating frame, we obtain for a forward propagating field
\begin{eqnarray}\label{Kz}
K_z &=& \sqrt{k'^2-{m^2\over R^2}} \approx k' - {1\over 2k'}{m^2\over R^2}  \nonumber \\
&\approx& k  - {m\Omega\over\omega}k - {1\over 2k}{m^2\over R^2}  .
\end{eqnarray}
In the last line the second term accounts for the rotational Doppler effect, and the third term accounts for the reduction in the z-component of the wavevector due to the skewing associated with the helical phase-front of the POV [17,18].  These terms have been kept only to leading order.

Based on the above results the slowly varying electric field envelope for a POV around the peak of the ring may be written as
\begin{equation}
A(\rho=R,\phi,z) = a(z)e^{im\phi}e^{-{iz\over 2k}{m^2\over R^2} -ikz{m\Omega\over \omega}}  .
\end{equation}
Then approximating the field of the POV as constant over its cross-section the power may be evaluated as
\begin{equation}\label{Ppov}
P(z)=2\epsilon_o nc \times 2\pi RW|a(z)|^2  .
\end{equation}
The utility of this solution rests on the Rayleigh range $z_R=kW^2/2$ being larger than the medium length $L$ so that the ring width will vary little under propagation through the medium.
\section{Appendix: Linearized signal-idler equations} 
The utility of the POV solution introduced in the main text rests on the Rayleigh range $z_R=kW^2/2$ being larger than the medium length $L$ so that the ring width $W$ will vary little under propagation through the medium.  Within this approximation the transverse Laplacian in cylindrical coordinates becomes $\nabla_\perp^2\rightarrow {1\over R^2}{\partial^2\over \partial\phi^2}$, thereby neglecting radial expansion of the ring.  Then treating the POV field in the vicinity of the ring radius $R$ as almost constant over the cross section, $A_j(\rho=R,\phi,z)\approx A_j(\phi,z)$, the field Eqs. (4) from the main text may be written along the top of the ring as
\begin{eqnarray}\label{A1}
\frac{\partial A_1}{\partial z}&-&{i\over 2k_1 R^2} {\partial^2 A_1\over\partial\phi^2} -k_1\left ({\Omega\over \omega_1} \right ){\partial A_1\over\partial \phi} + i\eta A_2A_{1}^*  =0 ,\nonumber \\
\frac{\partial A_2}{\partial z}&-&{i\over 4k_1 R^2}{\partial^2 A_2\over\partial\phi^2} - k_2\left ({\Omega\over \omega_2} \right ) {\partial A_2\over\partial \phi} =0 ,
\end{eqnarray}
where we have used the undepleted pump beam approximation in the lower equation.  Next the POV solution in Eqs. (6) is used to express the fundamental and second-harmonic fields as
\begin{eqnarray}\label{A2}
A_1(\phi,z) &=& a_1(z)e^{im_1\phi}e^{-{iz\over 2k_1}{m_1^2\over R^2} -ik_1z{m_1\Omega\over \omega_1}}+ \nonumber\\
&& a_3(z)e^{im_3\phi}e^{-{iz\over 2k_1}{m_3^2\over R^2} - ik_1z{m_3\Omega\over \omega_1}} , \nonumber \\
A_2(\phi,z) &=& a_2e^{im_2\phi}e^{-{iz\over 4k_1}{m_2^2\over R^2} - ik_2 z{m_2\Omega\over \omega_2}}.
\end{eqnarray}
Substituting Eqs. (\ref{A2}) into Eqs. (\ref{A1}) and using $m_3=m_2-m_1$ yields
\begin{eqnarray}
& & e^{im_1\phi}e^{-{iz\over 2k_1}{m_1^2\over R^2} -ik_1z{m_1\Omega\over \omega_1}}\left [ 
{da_1\over dz}-i(\eta a_2) a_3^*e^{i\kappa z} \right ] \nonumber \\
&+& e^{im_3\phi}e^{-{iz\over 2k_1}{m_3^2\over R^2} - ik_1z{m_3\Omega\over \omega_1}}\left [ 
{da_3\over dz} -i(\eta a_2) a_1^*e^{i\kappa z}
\right ] = 0  ,
\end{eqnarray}
with $\kappa$ given by Eq. (8) of the main text.  Then setting terms with the same azimuthal variation individually to zero, assuming $m_1\ne m_3$, yields the linearized signal-idler Eqs. (7) of the main text.
\section{Appendix: Parametric interaction of POVs}
Equations (7) of the main text may be solved for the fields at the output of the crystal of length $L$ [13,14]
\begin{eqnarray}
a_1(L) &=& a_1(0)\left ( \cosh(gL) - {i\kappa\over 2g}\sinh(gL)\right )e^{i\kappa L/2}, \nonumber \\
a_3(L) &=& a_1^*(0)\left ({i\eta a_2\over g}\right )\sinh(gL)e^{i\kappa L/2} ,
\end{eqnarray}
where $g=\sqrt{\eta^2|a_2|^2-\kappa^2/4}$ is the growth rate if the argument of the square root is positive. The field intensities are given by $I_j(z)=2\epsilon_o nc |a_j(z)|^2$ in terms of which the growth rate may be written as
\begin{equation}\label{g}
g = \sqrt{\beta I_p -\kappa^2/4} ,
\end{equation}
with $I_p=I_2(0)$ is the pump intensity at the peak of the ring and $\beta=(2\omega_1^2d_{eff}^2/\epsilon_0n^3c^3)$.  Using Eq. (\ref{Ppov}) the input signal power is $P_{sig}=2\epsilon_o nc \times 2\pi RW|a_1(0)|^2$, the output fundamental power is $P_1(L)=2\epsilon_o nc \times 2\pi RW(|a_1(L)|^2+|a_3(L)|^2)$, and the output signal power is $P_s(L)=2\epsilon_o nc \times 2\pi RW|a_1(L)|^2$.  Then the net gain for the fundamental field may be expressed as  
\begin{eqnarray}\label{G}
G=\frac{P_1(L)}{P_{sig}}&=&\left | \cosh(gL) - {i\kappa\over 2g}\sinh(gL)\right |^2  \nonumber \\ &+& \left | {\eta a_2\over g}\sinh(gL)\right |^2,
\end{eqnarray}
and the signal gain becomes
\begin{equation}\label{Gs}
G_s=\frac{P_s(L)}{P_{sig}}=\left | \cosh(gL) - {i\kappa\over 2g}\sinh(gL)\right |^2  .
\end{equation}


\begin{thebibliography}{99}

\bibitem{Birula}  I. Bialynicki-Birula, Z. Bialynicka-Birula, Phys. Rev. Lett. {\bf 78}, 2539 (1997).
\bibitem{Padgett}  M. P. J. Lavery, F. C. Speirits, S. M. Barnett, M. J. Padgett, Science {\bf 341}, 537 (2013).
\bibitem{SpeLavPad14} F. C. Speirtis, M. P. J. Lavery, M. J. Padgett, and S. M. Barnett, {\sl Opt. Lett.} {\bf 39}, 2944 (2014).

\bibitem{magnetic} J. Otterbach, J. Ruseckas, R. G. Unanyan, G. Juzeliunas,  M. Fleischhauer, Phys. Rev. Lett. {\bf 104}, 033903 (2010).
\bibitem{Zhang} G. Li, T. Zentgraf, S. Zhang, Nature Phys. {\bf 12}, 736 (2016).

\bibitem{Zel}
Ya. B. Zel'dovich, Pis'ma Zh. Eksp. Teor. Fiz. {\bf 14}, 270 (1971); Zh. Eksp. Teor. Fiz. {\bf 62}, 2076 (1972); [JETP Lett. {\bf 14}, 180 (1971)][Sov. Phys. JETP {\bf 35}, 1085 (1972)].

\bibitem{Zel2} Ya. B. Zel'dovich, L. V. Rozhanskii, A. A. Starobinskii, Izvestiya Vysshikh Uchebnykh Zavedenii, Radiofizika, {\bf 29}, I008-I016 (1986).

\bibitem{Penrose} R. Penrose, General Relativity and Gravitation, {\bf 34}, 1141 (2002) [reprinted from Rivista del Nuovo Cimento, Numero Speziale I, 257 (1969)].

\bibitem{bomb} W.H. Press, S.A.  Teukolsky,  Nature {\bf 238}, 211 (1972).

\bibitem{Cardoso1}  V. Cardoso, R. Brito, J. L. Rosa, Phys.Rev. D {\bf 91}, 124026 (2015) 

\bibitem{Cardoso2} ``Superradiance'', R. Brito, V. Cardoso, P. Piani, Springer (2015)

\bibitem{TLA}
This result follows from either the damped electron oscillator model or a two-level atom description, see for example, R. W. Boyd, {\it Nonlinear Optics}, 3rd Ed. (Academic, New York, 2008), Chap. 3.2.

\bibitem{BekSch98}
J. D. Beckenstein and M. Schiffer, { Phys. Rev. D} {\bf 58}, 064014 (1998).

\bibitem{MIT} M. F. Maghrebi, R. L. Jaffe, M. Kardar, Phys. Rev. Lett. {\bf 108}, 230403 (2012).

\bibitem{BeyRabin67}
P. P. Bey and H. Rabin, { Phys. Rev.} {\bf 162}, 794 (1967).

\bibitem{Boyd} R. W. Boyd, {\it Nonlinear Optics}, 3rd Ed. (Academic, New York, 2008), Chap. 2.

\bibitem{LowRogFac14}
J. Lowney, T. Roger, D. Faccio, and E. M. Wright, {Phys. Rev. A} {\bf 90}, 05328 (2014).

\bibitem{OstRicArr13}
A. S. Ostrovsky, C. Rickenstorff-Parrao, and V. Arrizon, {Opt. Lett.} {\bf 38}, 534 (2013).

\bibitem{CheMalAri13}
M. Chen {\it et. al}, {Opt. Lett.} {\bf 38}, 4919 (2013).

\bibitem{DhoSimPad96}
K. Dholakia, N. B. Simpson, M. J. Padgett, and L. Allen, { Phys. Rev. A} {\bf 54}, R3742 (1996).

\bibitem{RogHeiWri13}
T. Roger, J. F. Heitz, E. M. Wright, and D. Faccio, {Sci. Rep.} {\bf 3}, 3491 (2013).

\bibitem{BenBot98}
C. M. Bender and S. Boettcher, {Phys. Rev. Lett.} {\bf 80}, 5243 (1998).

\bibitem{Guo09}
A. Guo, G.J. Salamo, D. Duchesne, R. Morandotti, M. Volatier-Ravat, V. Aimez, G.A. Siviloglou and D.N. Christodoulides, {Phys. Rev. Lett.} {\bf 103}, 093902 (2009).

\bibitem{Christo} C. E. R{\"u}ter, K. G. Makris, R. El-Ganainy, D. N. Christodoulides, M. Segev, D. Kip, Nat. Phys. {\bf 6}, 192 (2010).

\bibitem{BenBerMan02}
C. M. Bender {\it et al.}, {J. Phys. A:Math. Gen.} {\bf 35}, L467 (2002).

\bibitem{Ant15}
D. A. Antonsosyan, { Opt. Lett.} {\bf 40}, 4575 (2015).

\bibitem{Silv14}
M. G. Silveirinha, {\sl Phys. Rev. A} {\bf 90}, 013842 (2014).

\bibitem{LanSil16}
S. Lanneb{\`e}re and M. G. Silveirinha, {\sl Phys. Rev. A} {\bf 94}, 033810 (2016).

\bibitem{Brill} C.Y. She, G.C. Herring, H. Moosm{\" u}ller, S. A. Lee, Phys. Rev. Lett. {\bf 51} 1648 (1983).

\bibitem{kishan}  Y. Arita, M. Mazilu, K. Dholakia, Nat. Commun. {\bf 4}, 2374 (2013).

\bibitem{Davis1} B. S. Davis, L. Kaplan, J. H. McGuire, J. Opt. {\bf 15} 035403  (2013).

\bibitem{Davis2}  L. Kaplan, J. H. McGuire, Phys. Rev. {\bf 92}, 032702 (2015).

\bibitem{Ncomms} C. T. Schmiegelow, J. Schulz, H. Kaufmann, T. Ruster, U. G. Poschinger, F. Schmidt-Kaler, Nat. Commun. {\bf 7}, 12998 (2016).

\bibitem{molina} G. Molina-Terriza, J. P. Torres, L. Torner, Nat. Phys. {\bf 3}, 305 (2007).

\bibitem{boyd1} S. Franke-Arnold, G. Gibson, R. W. Boyd, M. J. Padgett, Science {\bf 333}, 65 (2011).

\bibitem{boyd2} E. Wisniewski-Barker, G. M. Gibson, S. Franke-Arnold, R. W. Boyd, M.  J. Padgett, Opt. Express {\bf 22}, 11690 (2014).

\end{thebibliography}
\end{document}